\documentstyle[12pt]{article}

\newcommand{\EQ}{\begin{equation}}
\newcommand{\EN}{\end{equation}}

\renewcommand{\thefootnote}{\fnsymbol{footnote}}
\begin{document}
\topmargin 0pt
\oddsidemargin=-0.4truecm
\evensidemargin=-0.4truecm
\newpage
\setcounter{page}{0}
\begin{titlepage}
\begin{flushright}
SISSA 9/93/EP\\
IC/93/10
\end{flushright}
\begin{center}
{\large PONTECORVO'S ORIGINAL OSCILLATIONS REVISITED}
\end{center}
\vspace{0.2cm}
\begin{center}
{\large E.Kh. Akhmedov
\footnote{On leave from Kurchatov Institute of Atomic Energy, 123182 Moscow,
Russia. E-mail: \\ akhmedov@tsmi19.sissa.it, ~akhm@jbivn.kiae.su},
 ~S.T. Petcov
\footnote{Istituto Nazionale di Fisica Nucleare, Sezione di Trieste,
 Trieste, Italy}
\footnote{Permanent address: Institute of Nuclear Research and Nuclear
Energy, Bulgarian Academy of Sciences, BG-1784 Sofia, Bulgaria}\\}
{\em Scuola Internazionale Superiore di Studi Avanzati\\
Strada Costiera 11, I-34014 Trieste, Italy} \\
\vspace{0.4cm}
{\large and}\\
\vspace{0.4cm}
{\large A.Yu. Smirnov}
\footnote{On leave from Institute for Nuclear Research, Russian Academy of
Sciences, 117312 Moscow, Russia. E-mail: smirnov@trieste.ictp.it,
smirnov@inucres.msk.su}\\
{\em International Centre for Theoretical Physics, I-34100 Trieste, Italy}\\
\end{center}
\vspace{0.6cm}

\begin{abstract}
We show that a left-handed neutrino $\nu_L$ can oscillate into its $CP$-
conjugated state $\bar{\nu}_R$ with maximal amplitude, in direct analogy
with $K^0-\bar{K}^0$ oscillations. Peculiarities of such oscillations
under different conditions are studied.
\end{abstract}
\vspace{1cm}
\vspace{.5cm}
\end{titlepage}
\renewcommand{\thefootnote}{\arabic{footnote}}
\setcounter{footnote}{0}
\newpage
\section{Introduction}
35 years ago Pontecorvo suggested the possibility of neutrino oscillations
\cite{Pont1,Pont2} (see also \cite{Pont3})
by analogy with the oscillations of neutral $K$ mesons. The question he raised
was "...whether there exist other mixed neutral particles besides the $K^0$
mesons which differ from their antiparticles and for which the particle
$\to$ antiparticle transitions are not strictly forbidden" \cite{Pont1}.

Direct analogy with the $K^0-\bar{K}^0$ case would imply oscillations of a
neutrino into its $CP$-conjugated state with large amplitude. We shall refer
here to such a process as "Pontecorvo's original oscillations".

The essential difference between the $K^0$ mesons and neutrinos is related
to the spin of neutrinos. It was realized
after the $V-A$ structure of weak interactions had been established that
neutrinos are produced and interact in chiral states. In particular, only
left-handed neutrinos $\nu_L$ have been observed. The $CP$ conjugation
transforms $\nu_L$ into a right-handed antineutrino $\bar{\nu}_R$, and so
the realization of the Pontecorvo's original idea would mean the existence
of the oscillations
\EQ
\nu_L \leftrightarrow \bar{\nu}_R.
\EN
Strictly speaking, transitions (1) are not just a process of lepton
number oscillations, but also simultaneously neutrino spin precession.

Since the helicity of a free particle is conserved, in vacuum the oscillations
(1) cannot occur. Flavour oscillations between the
neutrinos of the same chirality are possible in vacuum \cite{Maki,Pont3}, but
in this case the transitions take place between the neutrino states which are
not related by $CP$ conjugation. In addition, the mixing of these states
need not be large.

Particle-antiparticle transitions in vacuum can in principle take place  for
4-component neutrinos. In terms of chiral states these oscillations would
imply transitions of $\nu_L$ into $\bar{\nu}_L$ (or $\bar{\nu}_R$ into
$\nu_R$) \cite{Pont2}, so that the neutrino helicity is conserved. However,
such transitions are not analogous to the $K^0-\bar{K}^0$ oscillations:
$\bar{\nu}_L$ is not the true antiparticle of the left-handed neutrino, the
existence of which is required by the $CPT$ invariance, but rather a different
neutrino state.
In the ultrarelativistic limit $\nu_L$ and $\bar{\nu}_L$ can be considered
as independent particles with quite different interactions ($\nu_L$ is
active whereas $\bar{\nu}_L$ is sterile in the standard model). This is in
contrast with the hypothetical neutron-antineutron oscillations, since at
low energies the different helicity components of the neutron are strongly
coupled through its large mass. Moreover, left-handed neutron and
antineutron have the same strong interaction.

For the above reasons it was generally supposed that Pontecorvo's
original oscillations are just the oscillations of active neutrinos into
sterile states, whereas the true neutrino-antineutrino oscillations (1) were
considered impossible. In this paper we show that under certain conditions
maximal-amplitude $\nu_L \leftrightarrow \bar{\nu}_R$ oscillations can
nevertheless occur.
\section{General conditions for $\nu_L \leftrightarrow \bar{\nu}_R$
transitions}
Consider for definiteness the transitions involving electron neutrinos,
$\nu_{eL} \leftrightarrow \bar{\nu}_{eR}$. As we already stressed,
oscillations (1) imply helicity flip of the neutrino states. Such a flip
can be induced, for example, by interactions
of neutrinos with external magnetic fields provided the neutrinos have magnetic
(or electric) dipole moments. However, the magnetic-moment interaction cannot
transform a neutrino into its own antineutrino because of $CPT$ invariance.
Nevertheless, it can convert a neutrino into an antineutrino of another species
\cite{ShVVVO}. From this fact two conclusions follow: (i) in addition to the
neutrino of a given flavour, one needs at least one more neutrino
state $\nu_x$ to be involved in the process, and (ii) an additional interaction
which mixes $\nu_x$ with $\nu_e$ is required. If these conditions are
satisfied, the $\nu_{eL} \rightarrow \bar{\nu}_{eR}$ transition
can proceed via $\nu_x$ in the intermediate state, and $\nu_{eL}-
\bar{\nu}_{eR}$ mixing appears as a second-order effect.

In the simplest case, the additional interaction should not
change the helicity of neutrinos, but it must change their lepton numbers.
Indeed, the $\nu_{lL}\leftrightarrow \bar{\nu}_{lR}$ oscillations
imply $\Delta L_l=2$ transitions, whereas in the magnetic-moment induced
transitions the individual lepton charges $L_l$ change by only one unit.
The additional interaction one needs can then be just the one which generates
the flavour (mass) mixing of neutrinos.

In all cases at least two chains of transitions contribute
to the $\nu_{eL}\leftrightarrow \bar{\nu}_{eR}$ transition, and one should
make sure that no cancellation between the corresponding amplitudes occurs.
Indeed, let $\nu_x$ be a left-handed neutrino; then the $\nu_{eL} \rightarrow
\bar{\nu}_{eR}$ transitions can proceed through the $\nu_{eL} \rightarrow
\nu_{xL} \rightarrow \bar{\nu}_{eR}$ chain, where the first transition is due
to the mass mixing and the second one is induced by the magnetic-moment
interaction. From $CPT$ invariance it follows that the antiparticle of
$\nu_{xL}$ exists, and is a right-handed neutrino $\bar{\nu}_{xR}$.
It also mediates the $\nu_{eL}
\rightarrow \bar{\nu}_{eR}$ transitions through the chain $\nu_{eL}
\rightarrow \bar{\nu}_{xR} \rightarrow \bar{\nu}_{eR}$. Now the first
transition is
due to the magnetic-moment interaction, and the second one results from the
mass mixing. The amplitudes of the helicity-flipping transitions $\nu_{eL}
\rightarrow \bar{\nu}_{xR}$ and $\nu_{xL} \rightarrow \bar{\nu}_{eR}$
in these two chains have opposite signs due to the $CPT$ symmetry
\footnote{The matrix of transition magnetic moments is antisymmetric.},
while the amplitudes of the transitions induced by the mass mixing coincide.
Therefore, if $\nu_{xL}$ and $\bar{\nu}_{xR}$ are degenerate in energy,
the contributions of the two chains to the $\nu_{eL} \rightarrow
\bar{\nu}_{eR}$ amplitude exactly cancel each other and the
$\nu_{eL}-\bar{\nu}_{eR}$ mixing does not appear. Obviously, the cancellation
takes place for any number of additional neutrinos. Moreover, this result
holds true for any number of the transitions in the chains: the crucial points
are that (i) there should be an
odd number of transitions induced by the magnetic-moment interaction, and
(ii) for a given chain, another one with $CP$-conjugated particles and
inverted order of transitions in the intermediate states always exists.

To induce the $\nu_{eL}\leftrightarrow \bar{\nu}_{eR}$ transitions, one should
lift the degeneracy of the intermediate states. This can be realized if the
transitions
take place in matter (provided the intermediate neutrino is not sterile)
and/or in a magnetic field whose direction changes along the neutrino
trajectory. Indeed, matter affects neutrinos and antineutrinos
differently (the corresponding forward scattering amplitudes
have opposite signs \cite{MSW}), and rotating magnetic fields affect
differently left-handed and right-handed states \cite{Sm,AKS}.

The possibility of $\nu_{eL}\leftrightarrow \bar{\nu}_{eR}$ transitions in
matter and transverse magnetic field was first pointed out in \cite{LM}.
However, for fixed-direction magnetic fields the transition probability
was shown to be small even for large neutrino mixing and magnetic moments
\cite{AKHM3}. As we shall see, the magnetic field rotation can change the
situation drastically.

The $\nu_{eL}\leftrightarrow \bar{\nu}_{eR}$ oscillations could in principle
be generated by the magnetic-moment interaction alone (i.e. could occur
even in the absence of mass mixing) if at least two additional neutrino states,
say, $\nu_{\mu}$ and $\nu_{\tau}$, and three transition magnetic moments
are involved. However, as it was shown in \cite{ASh2}, for massless
neutrinos the $\nu_{eL}-\bar{\nu}_{eR}$ mixing vanishes identically in this
case even in the presence of rotating magnetic field. The reason is
that equal numbers of left-handed and right-handed neutrinos are present
in the intermediate states, and so the cancellation of amplitudes is not
destroyed even by the field rotation. The same conclusion can be shown
to hold true also in matter since the properties of $\nu_{\mu}$ and
$\nu_{\tau}$ in matter are identical. This result changes
if neutrinos possess nonzero masses and vacuum mixing.

In what follows we shall discuss the $\nu_{eL}\leftrightarrow \bar{\nu}_{eR}$
oscillations induced by flavour mixing and transition magnetic moment with
one additional active neutrino, say, $\nu_{\mu}$. The case in which $\nu_x$
is a sterile neutrino will also be commented on.
\section{$\nu_{eL} \leftrightarrow \bar{\nu}_{eR}$ oscillations}
Consider a system of four neutrino states $\nu_{eL}$,
$\bar{\nu}_{eR}$, $\nu_{\mu L}$, and $\bar{\nu}_{\mu R}$ with vacuum mixing
and transition magnetic moment $\mu$ relating  $\nu_{eL}$ with
$\bar{\nu}_{\mu R}$. The evolution of this system in matter and magnetic
field can be described by the Schroedinger-like equation $i(d/dt)\nu=H\nu$,
where $\nu=(\nu_{eL},~\bar{\nu}_{eR},~\nu_{\mu L},~\bar{\nu}_{\mu R})^T$ and
$H$ is the effective Hamiltonian of the system:
\EQ
H~=~\left (
\begin{array}{cccc}
   0 & 0 & s_{2}\delta & \mu B_{\bot}\\
  0 & -2N-\dot{\phi} & -\mu B_{\bot} & s_{2}\delta\\
  s_{2}\delta & -\mu B_{\bot} & H_{\mu} & 0\\
  \mu B_{\bot} & s_{2}\delta & 0 & H_{\bar{\mu}}
\end{array}\right )
\EN
Here the angle $\phi (t)$ defines the direction of the magnetic field
${\bf B}_{\bot}(t)$ in the plane orthogonal to the neutrino momentum,
$B_{\bot}(t) = |{\bf B}_{\bot}(t)|$, ${\dot {\phi}}\equiv d\phi/dt$,
$N\equiv \sqrt{2}G_{F}(n_{e}-n_{n}/2),~~\delta\equiv \Delta m^{2}/4E$,
$s_2\equiv\sin 2\theta_{0},~~c_{2}\equiv \cos 2\theta_{0}$, where
$G_{F}$ is the Fermi constant, $n_{e}$ and $n_{n}$ are the electron
and neutron number densities, $E$ is the neutrino energy,
$\Delta m^{2}=m_{2}^{2}-m_{1}^{2}$, $m_1$, $m_2$ and $\theta_{0}$ being the
neutrino masses and mixing angle in vacuum. The matrix elements $H_{\mu}$ and
$H_{\bar{\mu}}$ in (2) read
\EQ
H_{\mu}\equiv -(1+r)N+2c_{2}\delta, ~~~H_{\bar{\mu}}
\equiv -(1-r)N+2c_{2}\delta-\dot{\phi}
\EN
where $r\equiv n_n/(2n_e-n_n)$. The diagonal elements of $H$ define the
energies of the "flavour levels", i.e. of $\nu_{eL}$, $\bar{\nu}_{eR}$,
$\nu_{\mu L}$ and $\bar{\nu}_{\mu R}$
\footnote{In deriving eq. (2) we have moved to the reference frame rotating
with the same angular velocity as the transverse magnetic field \cite{Sm}
and also subtracted a matrix proportional to the unit one so as to make
the first diagonal element equal to zero. These transformations amount to
multiplying neutrino states by certain phase factors and thus do not affect
the transition probabilities.}.
According to our previous discussion,
direct $\nu_{eL}-\bar{\nu}_{eR}$ mixing is absent in the Hamiltonian
(2), but can be induced in higher orders.

To have practically pure
$\nu_{eL}\leftrightarrow \bar{\nu}_{eR}$ transitions one should find the
conditions under which the ($\nu_{eL},\bar{\nu}_{eR}$) subsystem
approximately decouples
from the rest of the neutrino system. As we have indicated earlier,
this decoupling should not be complete, otherwise the effective
$\nu_{eL}-\bar{\nu}_{eR}$ mixing would disappear. We shall assume that the
following decoupling conditions are satisfied:
\EQ
|H_{\mu}|, |H_{\bar{\mu}}| \gg |2s_{2}\delta|, 2\mu B_{\bot}.
\EN
This allows  one to block-diagonalize the Hamiltonian
(2); the resulting effective Hamiltonian of the ($\nu_{eL},\bar{\nu}_{eR}$)
subsystem is
\EQ
H^{'}~=~\left ( \begin{array}{cc}
 0 & H_{e\bar{e}} \\
 H_{e\bar{e}}  &  -2N-\dot{\phi} + \eta
\end{array}\right )
\EN
Here the nondiagonal (mixing) matrix element is given by
\EQ
H_{e \bar{e}} = s_{2}\delta\,\mu B_{\bot}\,
\left(\frac{1}{H_{\mu}}-\frac{1}{H_{\bar{\mu}}}\right) \simeq
s_{2}\delta\,\mu B_{\bot}\,\frac{\dot{\phi}-2rN}
{[(-rN+\dot{\phi}/2)^{2}-(2c_2\delta)^{2}]}.
\EN
and $\eta \equiv (\tan \omega -\cot \omega)
H_{e\bar{e}}$, where
\EQ
\tan \omega \equiv s_{2}\delta/\mu B_{\bot}.
\EN
Note that the second equality in eq. (6) holds only for $|2N+\dot{\phi}|
\ll max\,\{(1+r)N,\,|2c_2\delta|\}$ (see the discussion below). It follows
from (6) that the effective $\nu_{eL}-\bar{\nu}_{eR}$ mixing is caused by
an interplay of flavour mixing and the one induced by the interaction of
the magnetic moment with magnetic field. In accordance with our general
discussion, it arises due to the transitions
through the $\nu_{\mu L}(\bar{\nu}_{\mu R})$ states: $\nu_{eL}
\rightarrow \nu_{\mu L}(\bar{\nu}_{\mu R})\rightarrow \bar{\nu}_{eR}$.
In fact, $H_{e\bar{e}}$ in eq. (6) exactly coincides with the result of the
calculations in the second-order perturbation theory, and the values
$H_{\mu}^{-1}$ and $H_{\bar{\mu}}^{-1}$ are just the propagators of the
Schroedinger equation corresponding to $\nu_{\mu L}$ and $\bar{\nu}_{\mu R}$
in the intermediate state. They enter eq. (6) with opposite signs
because of the antisymmetry of the matrix of transition magnetic moments.
In vacuum $H_{\mu}=H_{\bar{\mu}}$ and the contributions of
$\nu_{\mu}$ and $\bar{\nu}_{\mu }$ cancel each other. Matter ($n_n\ne 0$)
and magnetic field rotation ($\dot{\phi}\ne 0$) lift the degeneracy of the
$\nu_{\mu}$ and $\bar{\nu}_{\mu }$ levels and so give rise to the
$\nu_{eL}-\bar{\nu}_{eR}$ mixing.

It follows from eqs. (6) and (4) that the $\nu_{eL}-\bar{\nu}_{eR}$ mixing
term is always smaller than each of the generic first order mixings
$s_{2}\delta$ and $\mu B_{\bot}$. The better the decoupling, the smaller
$H_{e\bar{e}}$. The mixing term $H_{e\bar{e}}$ increases with decreasing
$H_{\mu}$ or $H_{\bar{\mu}}$, but this enhancement is limited by conditions
(4). The $\nu_{eL}-\bar{\nu}_{eR}$ mixing angle $\theta_m$ is defined as
\EQ
\tan 2\theta_m=\frac{2H_{e\bar{e}}}{-2N-\dot{\phi}+\eta}\;.
\EN

In medium with constant $N$, $r$, $B_{\bot}$ and $\dot{\phi}$,
the evolution of the $\nu_{eL}-\bar{\nu}_{eR}$ system will have a character
of oscillations with constant amplitude and length:
\EQ
P(\nu_{eL}\rightarrow \bar{\nu}_{eR};t)=\frac{(2H_{e\bar{e}})^2}
{(2H_{e\bar{e}})^2+(2N+\dot{\phi}-\eta)^2}\sin^2 \left(\frac{1}{2}
\sqrt{(2H_{e\bar{e}})^2+(2N+\dot{\phi}-\eta)^2}\,t\right )
\EN

\noindent Let us stress that in nonrotating magnetic fields the $\nu_{eL}-
\bar{\nu}_{eR}$
mixing is always strongly suppressed. Indeed, for $\dot{\phi}=0$ one gets
$\tan 2\theta_m \simeq 2r(\mu B_{\bot}\,s_2\delta)/(H_{\mu}H_{\bar{\mu}})\ll
1$.  On the contrary, in a twisting field the $\nu_{eL}\leftrightarrow
\bar{\nu}_{eR}$ oscillations can be {\em resonantly enhanced}. For
\EQ
\dot{\phi}=-2N+\eta \simeq -2N
\EN
the $\nu_{eL}-\bar{\nu}_{eR}$ mixing becomes maximal ($\sin^2 2\theta_m=1$)
and the oscillations proceed with maximal depth. Eq. (10) is nothing else but
the resonance condition for the $\nu_{eL}\leftrightarrow\bar{\nu}_{eR}$
oscillations. It implies that the field rotation compensates for the
energy splitting of the $\nu_{eL}$ and $\bar{\nu}_{eR}$ levels caused by
their interaction with matter. Up to the small term $\eta\sim H_{e\bar{e}}$,
it does not depend on the neutrino energy.

The necessity of matter and field rotation for strong $\nu_{eL}
\leftrightarrow \bar{\nu}_{eR}$ oscillations can therefore be understood as
follows. Matter is needed to avoid the cancellation of the
amplitudes with $\nu_{\mu}$ and $\bar{\nu}_{\mu}$ intermediate states in
(6). However, matter lifts the
degeneracy of the $\nu_{e}$ and $\bar{\nu}_{e}$ as well, and so the
$\nu_{eL}\leftrightarrow\bar{\nu}_{eL}$ oscillations will not proceed with
maximal amplitude. Field rotation can restore the degeneracy of $\nu_{e}$
and $\bar{\nu}_{e}$ while keeping the $\nu_{\mu}$ and $\bar{\nu}_{\mu}$
energies split, or vice versa. This comes about because the energies of the
electron and muon neutrinos in matter have different density dependence.
For the same reason the field rotation alone cannot lead to large-amplitude
$\nu_{eL}\leftrightarrow\bar{\nu}_{eR}$ oscillations.

The amplitude of the $\nu_{eL}\leftrightarrow\bar{\nu}_{eR}$ oscillations
in eq. (9) has a resonant dependence on the parameters of the problem. The
width of the resonant peak at half-height is
\EQ
\frac{\Delta \dot{\phi}}{|(\dot{\phi})_{res}|}=\frac{\Delta N}{N_{res}}=
\frac{2|H_{e\bar{e}}|}{N_{res}} \simeq 4(1+r)\,\frac{\mu B_{\bot}\,s_2\delta}
{|H_{\mu}\,H_{\bar{\mu}}|}  \ll 1,
\EN
i.e. the density and $\dot{\phi}$ widths of the resonance are very
small. On the contrary, due to the fact that the resonance condition is almost
energy-independent, the energy width of the resonant peak can be fairly large:
\EQ
\frac{\Delta E}{E_{res}}=\left \{\begin{array}{ll} 2(s_2\delta_0/
\mu B_{\bot})\cong
2\left [1-\frac{2(2N+\dot{\phi})c_{2}^2}{(1+r)N s_{2}^2 }\right ]^{-1/2},~&
(1+r)N \ll 2c_2\delta,\\
2(\mu B_{\bot}/s_2\delta_0)\cong
2\left[1-\frac{(1+r)N\,(2N+\dot{\phi})}{2(\mu B_{\bot})^2}\right ]^{-1/2},~&
(1+r)N \gg 2c_2\delta
\end{array} \right.
\EN
($\delta_0$ is the value of $\delta$ at resonance).
This means that for a neutrino beam with continuous energy spectrum a large
fraction of neutrinos can undergo resonantly enhanced $\nu_{eL}\leftrightarrow
\bar{\nu}_{eR}$ oscillations.

{}From eq. (9) it follows that the $\nu_{eL}\leftrightarrow\bar{\nu}_{eR}$
oscillation length at resonance $(l_{e\bar{e}})_{res}=\pi/H_{e\bar{e}}$ is much
bigger than the lengths of both the
flavour oscillations $\nu_{eL}\leftrightarrow \nu_{\mu L}$ and the spin-flavour
precession $\nu_{eL}\leftrightarrow\bar{\nu}_{\mu R}$ caused by the generic
first-order mixings. The latter two processes will proceed with small
amplitudes $\sim (s_2\delta/H_{\mu})^2$ and $(\mu B_{\bot}/H_{\bar{\mu}})^2$
respectively. Therefore the $\nu_e$ transition probability will be described
by a superposition of maximal-amplitude long-wavelength and small-amplitude
short-wavelength oscillations (Fig. 1a).
\section{$\nu_{eL} \leftrightarrow \bar{\nu}_{eR}$ oscillations influenced
by a third neutrino state}
If the flavour energy levels $H_{\mu}$ or $H_{\bar{\mu}}\to 0$, the $\nu_{e}-
\bar{\nu}_{e}$ mixing term
$H_{e\bar{e}}$ and the width of the $\nu_{e}-\bar{\nu}_{e}$ resonance
increase, but the influence of the $\nu_{\mu}$ or $\bar{\nu}_{\mu}$ levels on
the $\nu_{eL}\leftrightarrow \bar{\nu}_{eR}$ transition probability becomes
stronger. In particular, the amplitudes of the $\nu_{eL}\leftrightarrow
\nu_{\mu L}(\bar{\nu}_{\mu R})$ transitions increase.
In this case the $\nu_{eL}\leftrightarrow\bar{\nu}_{eR}$ resonance density
approaches the MSW one \cite{MSW} or that of the resonant spin-flavour
precession \cite{LM,AKHM1}.

Let us assume that
\EQ
\frac{\dot{\phi}}{2}=-N=-\frac{2c_2\delta}{1+r},
\EN

\noindent which corresponds to $H_{\mu}=0$. Now the energy levels of three
neutrino states, namely, those of $\nu_{eL}$,
$\bar{\nu}_{eR}$ and $\nu_{\mu L}$, cross in one point (this means that
the resonances of the $\nu_{eL}
\leftrightarrow\bar{\nu}_{eR}$, $\nu_{eL}\leftrightarrow \nu_{\mu L}$ and
$\nu_{\mu L}\leftrightarrow \bar{\nu}_{eR}$ transitions merge in one point).
The influence of the
$\nu_{\mu}$ state on the ($\nu_{eL},\bar{\nu}_{eR}$) system becomes maximal.
However, if $H_{\bar{\mu}}$ satisfies eq. (4) (which is realized for $s_2 \ll
1$ and $\mu B_{\bot} \ll 4c_2\delta$), the $\bar{\nu}_{\mu}$ state
will still be decoupled, and so one can
perform a $3-1$ block-diagonalization of $H$. The resulting effective
Hamiltonian of the strongly coupled ($\nu_{eL},\bar{\nu}_{eR},\nu_{\mu L}$)
system is just given by the $3\times 3$ matrix in the upper left corner of
$H$ in eq. (2) with the following modifications: (i) the diagonal terms
acquire small corrections which are not important for our consideration,
and (ii) the $H_{e\bar{e}}=H_{\bar{e}e}$ terms are no longer zero but rather
are equal to (${}-\epsilon$), where
\EQ
\epsilon = \frac {(s_{2}\delta)(\mu B_{\bot})}{H_{\bar {\mu}}}.
\EN
The eigenvalues of this Hamiltonian are
\EQ
H_{1} \cong f -\frac{\epsilon'}{2},~~
 \, \, H_{2} \cong - f -\frac{\epsilon'}{2}, \,
  \,~~H_{3} \cong \epsilon',
\EN
where
\EQ
f = \sqrt{(\mu B_{\bot})^{2} + (s_{2}\delta)^{2}},~~
\epsilon ' = - \epsilon \sin 2\omega,
\EN
and $\omega$ is defined by eq. (7). The orthogonal matrix diagonalizing
the effective Hamiltonian is given to leading order in $\epsilon/f$ by
\EQ
S_{m}
{}~=~ \left (
\begin{array}{ccc}
   \frac{\sin \omega}{\sqrt{2}}
 & \frac{\sin \omega}{\sqrt{2}} & \cos \omega \\
 -\frac {\cos \omega}{\sqrt{2}} &
             -\frac{\cos \omega}{\sqrt{2}} &  \sin \omega \\
\frac{1}{\sqrt{2}}  & - \frac{1}{\sqrt{2}}
    &  -\frac{\epsilon}{f}\, \cos 2\omega
\end{array}\right ).
\EN
For constant $N$, $r$, $B_{\bot}$ and $\dot{\phi}$ one obtains from
(15)--(17) the following probability of the $\nu_{eL}\leftrightarrow
\bar{\nu}_{eR}$ oscillations:
\EQ
P(\nu_{eL}\rightarrow\bar {\nu}_{eR};t) =
 \sin^{2} 2\omega \, \sin^{4} {\frac{1}{2}ft} +
\sin^{2} 2\omega \, \sin^{2} {\frac{3}{4} \epsilon' t} \,
  \cos ft +
  d \, \sin {\epsilon' t}\, \sin {ft},
\EN
where $d = O(\epsilon)$. The first term in (18) corresponds to the
limit $\epsilon\to 0$. We see that in the three-level crossing point the depth
of the $\nu_{eL}-\bar{\nu}_{eR}$ oscillations, $\sin^{2} 2\omega$, does not
exhibit
any suppression related to the higher order direct $\nu_{eL}-\bar {\nu}_{eR}$
mixing. Moreover, in the symmetric case $s_{2}\delta = \mu B_{\bot}$, when the
vacuum mixing is equal to that induced by the magnetic moment interaction, one
has $\sin 2\omega=1$ and the oscillation depth becomes maximal. For
$s_{2}\delta \neq \mu B_{\bot}$ the oscillation depth is less than unity
and it decreases when the difference between $s_{2}\delta$ and $\mu B_{\bot}$
increases.

The $\nu_{eL}\leftrightarrow\bar {\nu}_{eR}$ oscillation  length is given by
\EQ
l_{e\bar{e}} = \frac{2\pi}{f} =
    \frac{2\pi}{\sqrt{(s_{2}\delta)^{2} + (\mu B_{\bot})^{2}}}.
\EN
For $s_{2}\delta= \mu B_{\bot}$ it is only $\sqrt{2}$ times bigger than
flavour oscillation or spin-flavour precession lengths at the resonance point:
$l_{e\bar{e}} = \sqrt{2} l_{osc} = \sqrt{2} l_{p}$ ($l_{osc} =
4\pi E/(\Delta m^2 \sin 2\theta_0),~l_{p}=\pi/\mu B_{\bot}$). In contrast with
the case of pure $\nu_{eL}
\leftrightarrow\bar{\nu}_{eR}$ oscillations discussed above, $l_{e\bar{e}}$
does not contain any big factor like $f/\epsilon$; $P(\nu_{eL}\rightarrow
\bar{\nu}_{eR})$ depends on the fourth power of $\sin (\pi t/l_{e\bar{e}})$
rather than on the second power (see Fig. 1b). These features are related to
the fact that there are three levels involved and that the splitting between
the levels is determined by $f$. Now three
neutrino species oscillate into each other with comparable
amplitudes. For example, the probability of the $\nu_{eL}\rightarrow
\nu_{\mu L}$ oscillations is $P(\nu_{eL}\rightarrow \nu_{\mu L}; t) \cong
\sin^2 \omega\cdot \sin^2 ft$. Note that the oscillation length for this
mode is two times as small as $l_{e\bar{e}}$. For maximal depth of the
$\nu_{eL}\leftrightarrow\bar{\nu}_{eR}$ oscillations the amplitude of the
$\nu_{eL}\leftrightarrow \nu_{\mu L}$ oscillations is $\sin^2 \omega=1/2$.

The second term in (18) is generated by the direct $\nu_{eL}-\bar {\nu}_{eR}$
mixing $\sim \epsilon$ in the $3\times3$ effective Hamiltonian. It gives a
long-period modulation of the oscillation probability. The corresponding
modulation length is $l_{mod} \cong 4\pi/(3\epsilon') \gg l_{e\bar{e}}$.
The first two terms in (18) can also be written as $\frac{1}{4}\sin^{2}
2\omega \,(1+\cos^{2} ft - 2\cos ft \, \cos \frac{3}{2} \epsilon't)$,
which implies that the modulation leads to the oscillation depth varying
between $\sin^{2} 2\omega$ and $\frac{1}{2}\sin^{2} 2\omega$ (Fig. 1b).

Let us stress that the maximal-amplitude short-wavelength oscillations
described by eq. (18)
are only possible in a rotating field, when the merging condition (13)
can be fulfilled. The merging condition and $l_{e\bar{e}}$ depend on the
neutrino energy, and so the enhancement of the $\nu_{eL}\leftrightarrow
\bar{\nu}_{eR}$ oscillations in this case has a resonant character, too.

\section{Discussion and conclusions}
We have discussed so far $\nu_{eL}\leftrightarrow \bar{\nu}_{eR}$
transitions with an active neutrino in the intermediate state. Evidently,
instead of $\nu_{\mu}$ or $\nu_{\tau}$, a sterile neutrino could play the role
of the additional neutrino required for the $\nu_{eL}\leftrightarrow
\bar{\nu}_{eR}$ oscillations. All the above results hold in this
case as well; the corresponding analytical expressions can be obtained from
those already derived by setting $r=0$. Now, according to
(6), $H_{e\bar{e}}\sim \dot{\phi}$ and the $\nu_{eL}\leftrightarrow
\bar{\nu}_{eR}$ transitions can only occur in twisting magnetic fields:
matter alone is not sufficient to induce the $\nu_{eL}\leftrightarrow
\bar{\nu}_{eR}$ oscillations. Sterile neutrinos do not interact with matter
and so the degeneracy of $\nu_{xL}$ and $\bar{\nu}_{xR}$ can only be lifted
by the magnetic field rotation.

Our previous discussion was mainly constrained to the case of constant
$N$, $r$, $B_{\bot}$ and $\dot{\phi}$. In a medium with matter density,
and magnetic field varying along the neutrino path,
the maximum-amplitude $\nu_{eL}\leftrightarrow\bar{\nu}_{eR}$
oscillations will take place if the functions $N(t)$ and $\phi(t)$ satisfy
the resonance condition (10) over a sufficiently large space interval
$\Delta t$.

Eq. (10) is in fact just the condition of crossing of the $\nu_{e}$ and
$\bar{\nu}_{e}$ levels. Thus, if the parameters of the medium vary in such
a way that at a certain point $t_r$ eq. (10) holds, resonant
$\nu_{eL}\rightarrow\bar{\nu}_{eR}$ conversion may take place in the
resonant region \cite{APS}. The conversion can be nearly complete if the
matter density and $\dot{\phi}$ vary slowly enough (adiabatically) along the
neutrino path.

The maximum-amplitude $\nu_{eL}\leftrightarrow\bar{\nu}_{eR}$ oscillations
we have discussed, especially those in the merging point, can provide an
efficient mechanism of generation of $\bar{\nu}_{e}$ flux in the sun.
They can also have important consequences for neutrinos created in
collapsing stars.

In conclusion, we have shown that the original Pontecorvo's idea of
large-amplitude neutrino-antineutrino oscillations can be realized provided
the following conditions are satisfied:

\noindent (i) at least one additional neutrino $\nu_x$ is involved in the
transitions, which is mixed with the initial neutrino $\nu_l$ through flavour
(mass) mixing; \\
(ii) there exists a transition magnetic moment which connects
the $\nu_l$ with $\bar{\nu}_x$;\\
(iii) neutrino propagates in matter and transverse magnetic field, and \\
(iv) the direction of magnetic field changes along the neutrino path.

There are two distinct cases of large-amplitude $\nu_{eL}\leftrightarrow
\bar{\nu}_{eR}$ oscillations: (a) practically pure $\nu_{eL}\leftrightarrow
\bar{\nu}_{eR}$ oscillations when $\nu_{eL}$ and $\bar{\nu}_{eR}$ decouple
from the rest of the system (the resonant condition (10) should
be satisfied); (b) $\nu_{eL}\leftrightarrow\bar{\nu}_{eR}$ oscillations in the
presence of strong transitions into a third neutrino. In this case the
merging condition (13) should be fulfilled and, in addition, the flavour
mixing should be approximately equal to the magnetic-moment induced one
($\mu B_{\bot}\simeq s_2\delta$).
The two cases are characterized by different oscillation lengths and
different forms of dependences of the oscillation probability on the
distance. \\

A.Yu.S. would like to thank Prof. A. Salam,
the International Atomic Energy Agency and UNESCO for hospitality at
the International Centre for Theoretical Physics.
The work of S.T.P. was supported in part by
the Bulgarian National Science Foundation via grant PH-16.

\newpage
\noindent {\Large Figure caption}

\vspace{1.0cm}
\noindent Fig. 1. Dependence of the $\nu_{eL}\leftrightarrow\bar{\nu}_{eR}$
oscillation probabilities on the distance travelled by neutrinos: (a) the
case of isolated $\nu_{eL}-\bar{\nu}_{eR}$ system, (b) $\nu_{eL}
\leftrightarrow\bar{\nu}_{eR}$ oscillations in the presence of strong
transitions into a third neutrino state.
\end{document}